\documentclass[runningheads]{llncs}
\usepackage{graphicx}
\usepackage{wrapfig}
\usepackage{color}
\newcommand{\raga}[1]{{\color{black}{#1}}} 
\newcommand{\cmn}[1]{{\color{black}{#1}}} 
\newcommand{\kty}[1]{{\color{black}{#1}}} 

\begin{document}
\title{Feasibility of Smartphone Vibrations\\ as a Sensory Diagnostic Tool\thanks{This work was supported in part by the Stanford Precision Health and Integrated Diagnostics Center, National Science Foundation Graduate Research Fellowships, Stanford Graduate Fellowships, and National Science Foundation grant 1830163. \textsuperscript{$\star\star$}These authors contributed equally to this work.}}
%
%
\author{Rachel A.\ G.\ Adenekan\inst{1}\and
Alexis \raga{J.\ }Lowber\inst{1}\and
Bryce \raga{N.\ }Huerta\inst{1} \and \\ 
Allison M.\ Okamura\inst{1}\and
Kyle T.\ Yoshida\inst{1\textsuperscript{$\star\star$}} \and
Cara M.\ Nunez\inst{1,2,3\textsuperscript{$\star\star$}}\
}


%
\authorrunning{R.\ \cmn{A. G. }Adenekan et al.}
%
\institute{Stanford University, Stanford, CA 94305, USA\\ \email{adenekan@stanford.edu} \and
Harvard University, Cambridge, MA 02138, USA
\and
Cornell University, Ithaca, NY 14850, USA
}

\maketitle              
\begin{abstract}
Traditionally, clinicians use tuning forks as a binary measure to assess vibrotactile sensory perception. This approach has low measurement resolution, and the vibrations are highly variable. Therefore, we propose using vibrations from a smartphone to deliver a consistent and precise sensory test. First, we demonstrate that a smartphone has more consistent vibrations compared to a tuning fork. Then we develop an app and conduct a validation study to show that the smartphone can precisely measure a user's absolute threshold. This finding motivates future work to use smartphones to assess vibrotactile perception, allowing for increased monitoring and widespread accessibility. 


\end{abstract}
\section{INTRODUCTION}
 Clinical tuning forks are commonly used to diagnose diminished vibrotactile sensory function and monitor changes over time. This method requires a clinician to manually conduct a vibration sensitivity test (VST) during which the clinician strikes a tuning fork, places it on the patient's skin, then asks the patient to verbally indicate if vibrations are perceived. \cmn{The highly variable tuning fork vibrations\raga{ and the use of only binary responses to a single vibration stimulus} leads to an imprecise VST.}
 
Due to the ubiquity of haptic actuators in mobile phones, vibrotactile perception can be tested outside of the lab or clinic~\cite{blum2019getting}. Smartphones have been used for sensory diagnostics~\cite{jasmin2021validity,may2017mobile}, but prior studies suffer from lack of characterization of the vibration stimulus, confounding factors, and use of only a binary measurement (similar to the tuning fork VST). We intend to mitigate these issues and assess the reliability of mobile phones as a research and diagnostic tool. Specifically, we aim to develop a \raga{non-binary} smartphone-based VST that enhances the precision \cmn{and accessibility} of diagnostic exams for tactile deficits.
\section{INSTRUMENT VIBRATION CHARACTERIZATION}
\begin{figure}
\begin{center}
\includegraphics[width=\textwidth]{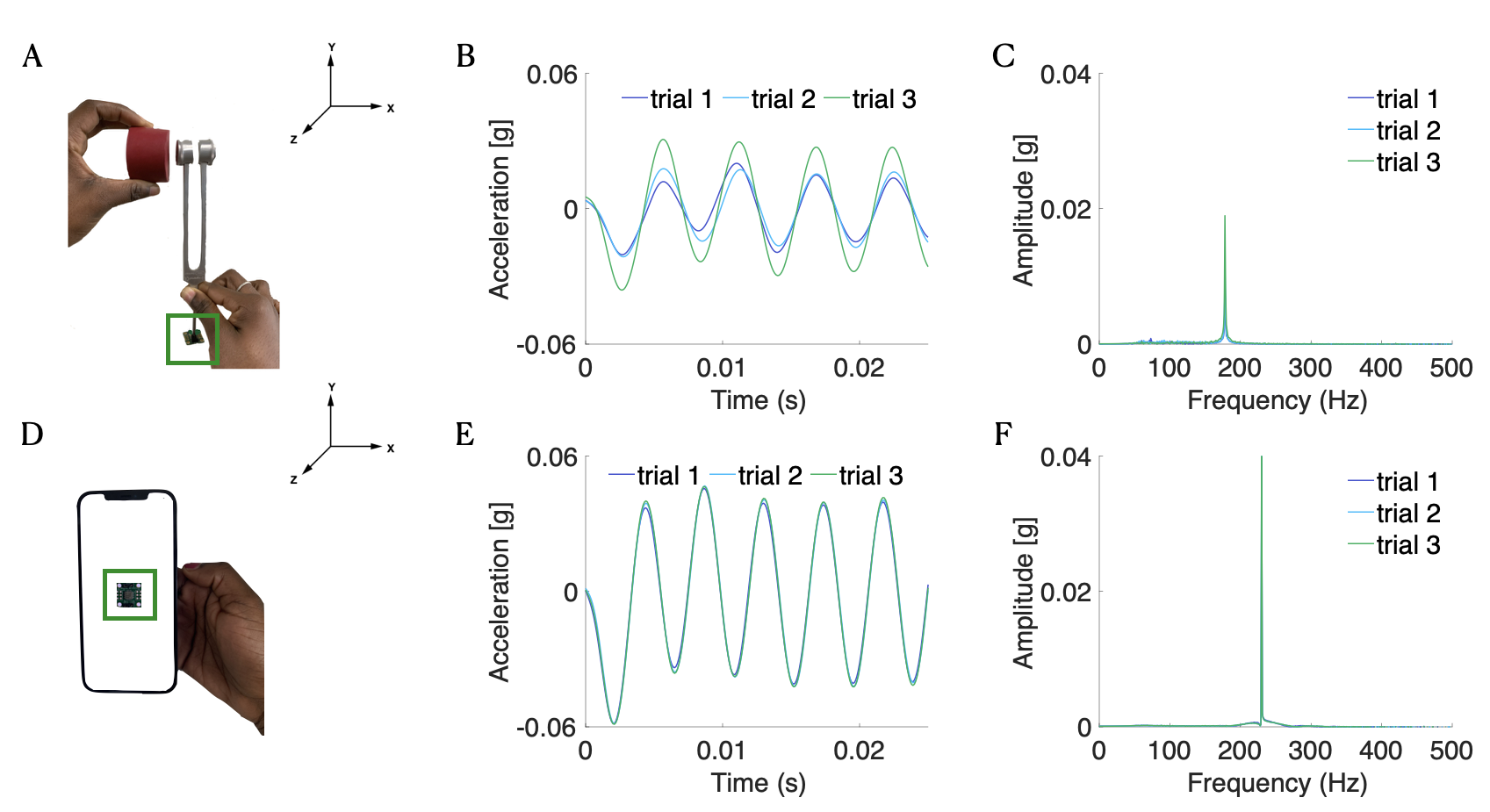}
\end{center}
\caption{\label{fig:fig1}Vibration measurement setup and data. A: 128 Hz tuning fork accelerometer placement (green box). B: Tuning fork filtered vibration waveforms. C: Tuning fork FFT for each trial (peak frequency = 178 Hz with varying \raga{amplitudes}). D: iPhone 12 Pro Max accelerometer placement (green box). E: iPhone filtered vibration waveforms. F: iPhone FFT for each trial (peak frequency = 230 Hz with similar \raga{amplitudes}).}
\end{figure}

To measure tuning fork vibrations present during a VST, we attached an accelerometer (Analog Devices, EVAL-ADXL354CZ, 3-axis $\pm2g$) to the base of a 128 Hz tuning fork (CynaMed). Then, we measured acceleration using a DAQ (National Instruments, NI9220), interfaced with MATLAB (Mathworks) (Fig~\ref{fig:fig1}A). The resulting vibration amplitudes varied between trials (Fig~\ref{fig:fig1}B). This aligns with a previous finding that tuning fork vibration waveforms are sensitive to the strength of the blow used to generate the vibrations~\cite{rossing1992acoustics}. We also found that the 128 Hz tuning fork \raga{resonated at 178 Hz instead of 128 Hz} (Fig~\ref{fig:fig1}C).

We also measured vibrations on the front-center of an Apple iPhone 12 Pro Max with the same accelerometer (Fig~\ref{fig:fig1}D). Using Apple's Core Haptics framework, continuous vibrations were delivered with a ``hapticSharpness" value of 1.0 and ``hapticIntensity" value of \raga{0.25} ($n=3$). Vibration acceleration waveforms (Fig~\ref{fig:fig1}E) and FFTs (Fig~\ref{fig:fig1}F) indicate that the smartphone can relay more consistent vibration amplitudes than the tuning fork. The iPhone had a peak frequency of 230 Hz, which is slightly higher than the tuning fork, but still in the range that stimulates the Pacinian corpuscles which respond to vibration~\cite{johnson1999vibration}. 

\section{PRELIMINARY TOOL VALIDATION}
We developed an iOS app that controls Apple Core Haptics variables and implements a staircase method (reversals = 8) to determine absolute \cmn{intensity} threshold. The ``hapticIntensity" varies with a step size of 0.05, while the ``duration" (0.1 seconds) and ``hapticSharpness" (1.0) are held constant. Time intervals between vibrations are randomized to reduce bias, and response times greater than 1.5 seconds are counted as false positives. Absolute intensity threshold is calculated by averaging the vibration intensity readings at the reversal indices.

To test the precision of \kty{our tool}, we conducted ten trials with the app on one healthy participant in a single day. For consistency between trials, we instructed the participant to hold the phone such that all four fingertips are in contact with the back of the phone and to use the thumb to provide responses via a button within the app. Physiologically, we expect absolute \cmn{intensity} threshold to remain stable during this time period. For this participant, we calculated an absolute intensity threshold of 0.348 ($\pm0.040$)\cmn{. Since the standard deviation} is less than the ``hapticIntensity" step size of 0.05\cmn{, w}e conclude that our approach can \cmn{consistently and precisely} measure a participant's absolute intensity threshold. 

\section{CONCLUSIONS AND FUTURE WORK}
We demonstrate that smartphone-based vibrations can be used for a reliable, mobile VST. Next, we will quantify the clinical significance of our app's absolute intensity threshold by correlating it to clinical benchmarks, including the 128 Hz tuning fork and the Semmes-Weinstein monofilament exam, for both healthy users and patients with sensory neuropathy \raga{($n>20$)}. Lastly, we \raga{will} complete a comprehensive mechanical characterization of smartphone vibrations and refine our app so that it can be widely distributed as an in-home diagnostic tool.

%
%
\bibliographystyle{splncs04}
\bibliography{mybibliography} 
\end{document}